\documentclass[reprint,amsmath,amssymb,aps]{revtex4-2}
\usepackage{graphicx}
\graphicspath{ {images/} }
\usepackage{epstopdf}
\usepackage{hyperref}

\usepackage{color}
\definecolor{green}{rgb}{0,0.5,0}
\usepackage{textcomp}

\def\beq{\begin{eqnarray}}
\def\eeq{\end{eqnarray}}
\def\be{\begin{equation}}
\def\ee{\end{equation}}
\def\eq{&=&}

\def\bm{\begin{math}}
\def\me{\end{math}}

\def\la{\langle}
\def\ra{\rangle}

\def\del{\partial}

\def\bel{\begin{equation} \label}
\def\beel{\begin{eqnarray} \label}

\newcommand \bei {\begin{itemize}}
\newcommand \eei  {\end{itemize}}

\begin{document}


\title{ Aggregation kinetics at sedimentation: the impact of particles diffusion }

\author{R. Zagidullin$^{1,2}$}
\author{A.P. Smirnov$^{2}$}
\author{S. Matveev$^{2,3}$}
\author{N. V. Brilliantov$^{1,4}$}

\affiliation{$^{1}$Skolkovo Institute of Science and Technology, Moscow, Russia}
\affiliation{$^{2}$Faculty of Computational Mathematics and Cybernetics, Lomonosov Moscow State University, Russia}
\affiliation{$^{3}$Marchuk Institute of Numerical Mathematics of Russian Academy of Sciences, Moscow, Russia}
\affiliation{$^{4}$University of Leicester, Leicester, UK}


\begin{abstract}
We investigate the aggregation  kinetics of sedimenting particles theoretically and numerically, using the advection-diffusion equation. Agglomeration, caused by both transport mechanisms (diffusion and advection), is  important for small particles, like primary ash or soot particles in atmosphere, and large particles of equal or close size, where the advection mechanism is weak. For small Peclet numbers, which quantify the relative importance of diffusion and advection, we obtain the aggregation rates, as an expansion in Peclet numbers.  For large Peclet numbers we use purely ballistic aggregation rates. Combining these results we obtain the rational approximant for the whole range of Peclet numbers. We also compute the aggregation rates by numerically solving the advection-diffusion equation.  The  results of the numerical simulations are in excellent agreement with the analytical theory for the studied Peclet numbers, varying by  four orders of magnitude. 
\end{abstract}

\maketitle


\section{Introduction}
Sedimentation of coagulating particles is ubiquitous; it occurs in numerous natural and industrial processes, see e.g. \cite{cohes_sedim_JFM2019, agg_Brown_JCP2019,Langmuir2016,sedim_adhes_PT2021, sed_coll_JCP_2011}. Aggregation of dust, ash or soot falling in the air, coagulating organic particles sedimenting in water (lakes, rivers, seas) may be mentioned as  prominent examples. Particles falling in fluid  are subjected to  the driving force (gravity and buoyancy) and friction force. After a short transient period these forces equilibrate giving rise  to a steady velocity of a falling  particle. Different steady velocities result in collision of particles if their trajectories intersect. If a particle is small enough, a stochastic force originating from the molecular fluctuations in surrounding fluid becomes important. It gives rise to a random, diffusive component of a particle motion. For instance, the diffusion motion is comparable with the ballistic one for particles smaller than $2 \mu$ \cite{Pinsky, khodzher2021study, zhamsueva2022studies} in air, which corresponds to primary  soot particles~\cite{SHAHAD1989141}. 

Similarly, the diffusion of particles may cause collisions.  If the particles are cohesive (which is always the case for small particles, due to molecular surface forces) they stick together, forming a joint aggregate. These aggregates can further collide with other aggregates or particles. As the system evolves, larger and larger clusters emerge and the size distribution of the aggregates in the system permanently widens. We will characterize the aggregates, comprising $k$ particles (monomers), by their concentration $n_k(t)$ and radius $R_k$, assuming that they are spherical. Correspondingly, the concentration of primary particles (monomers) of radius $R_1$ is $n_1(t)$. 

The rate of the coagulation process, that is, the number of the aggregative collisions per unit time and unit volume of the aggregates of size $i$ and $j$ may be written as $K_{ij}n_in_j$ \cite{krapbook,Leyvraz2003,Smo17}. Namely, it is proportional to the concentration of particles $n_i$ and $n_j$ and the rate  constant $K_{ij}$ which depends on the nature of the particles, their size, and the specific aggregation process. In the classical Smoluchowski case of diffusing particles, uniformly distributed in space and in the lack of external forces, these constants read, $K_{ij}=4\pi (D_i+D_j)(R_i+R_j)$, where $D_i$ and $D_j$ are respectively the diffusion coefficients of the aggregates $i$ and $j$ \cite{krapbook,Leyvraz2003,Smo17}. Hence the aggregation is driven by the mutual diffusion of the particles. 

If all aggregates move with the same steady velocity, the kinetic constants, naturally, will not change. However, the difference in the steady velocities changes the aggregation mechanism as the particles can collide by overtaking  each other. If particles are large enough, so that the diffusive motion may be neglected,   the rate constants follow from the purely ballistic trajectories of particles yielding $K_{ij}^{\rm bal}=\pi (R_i+R_j)^2 |v_i-v_j|$, where $v_i$ and $v_j$ are aggregate steady velocities, which corresponds to the Smoluchowski aggregation rates for a shear flow  \cite{JCEJ,SaffmanTurner}. To account for the hydrodynamic interactions between falling particles these rates are modified by using the collision efficiency factor $E(R_i, R_j)$ yielding  \cite{Pinsky,FalkovichNature,Falkovich2006}, 
\begin{equation}
 \label{KE}
  K_{ij}^{\rm bal}=\pi (R_i+R_j)^2 E(R_i, R_j)|v_i-v_j| .  
\end{equation}
The factor $E(R_i, R_j)$ may be obtained by the numerical solution of a complicated hydrodynamic problem for two spheres in a flux. Unfortunately, an explicit expression for this factor is presently lacking; it is available in a form of tables with the values of $E$ for different particle sizes and air pressure \cite{Pinsky}. 

For the case of sedimenting small particles both mechanisms -- the diffusional and due  to the velocity difference are important. Moreover,
even for large particles Eq. \eqref{KE} predicts vanishing aggregation rates for same-size particles for which steady velocities are equal. For close-size particles the rates are also small, so that the diffusion mechanism may not be neglected. 
The relative importance of the advection motion and diffusional motion  is quantified by the Peclet number $\rm Pe$ (see the definiton below). 
In Refs. \cite{Ven_and_Mason_1977,MELIK198484} the aggregation kinetics of sedimenting particles with  the strong prevalence of the diffusional component (small values of $\rm Pe$) has been studied analytically, while in Ref. \cite{Feke} the other limit of the strong prevalence of the advective motion (large values of $\rm Pe$) has been addressed. In Ref. \cite{sedim_adhes_PT2021} extensive numerical simulations by the lattice Boltzmann approach  in combination with the large eddies method have been performed. In all these studies however an  analytical expression for the rate kernel $K_{ij}$, adequate for the whole range of $\rm Pe$,  has not been provided. 

The goal of the present study is to construct explicit expressions for the reaction rate coefficients $K_{ij}$, describing aggregation of sedimenting particles, which may be used for arbitrary values of  Peclet number. Such  rate coefficients are particular important for the agglomeration process, when the value of $\rm Pe$ are very small for the primary particles and become large for mature aggregates. To obtain these expressions we consider  separately two limiting cases of small and large $\rm Pe$.  Then we construct a simple rational approximation which describes accurately the aggregation kinetics for the whole diapason of Peclet number. In this approach we neglect the hydrodynamic interactions between particles; to account for the latter interactions, we propose a simple generalization of our approximanion.  

The article is organized as follows. In the next Sec. II we develop a theoretical approach to the problem, while  Sec. III is devoted to its numerical analysis; here we also compare the results of the theory and simulations. Finally in Sec. IV we summarize our findings. 

\section{Aggregation  model of sedimenting particles }
Here we consider the sedimentation of particles at small Reynolds numbers which implies the validity of the linear hydrodynamic theory. The problem of multi-particle sedimentation is very complicated, therefore we assume that the system is very dilute and only pairwise  interactions between neighboring  particle are important; we also ignore the hydrodynamic interactions between the particles. 

Consider first sedimenting  single particle of mass $m$, radius $R$ and density $\rho_p$. If the particle is small enough it performs a random motion which obeys the Langevin equation: 

\bel{langeven}
m \frac{d {\bf v}}{dt}  = - \gamma {\bf v} + {\bf F}_{\rm b} +{\bf F}_{\rm st}.\ee
Here the viscous  friction force is  $-\gamma {\bf v}$ with $\gamma =6\pi \eta R$, where $\eta$ is the fluid viscosity. It acts along with the force ${\bf F}_b$ which combines the gravity and buoyancy $(4/3)\pi R^3(\rho_p-\rho_f)g$, where $\rho_f$ is the fluid density and $g$ is the gravitation acceleration. The stochastic force ${\bf F}_{\rm st}(t)$ has zero mean and the variance according to the dissipation-fluctuation theorem (FDT) \cite{krapbook}:
\beel{FDT1}
&&\la {\bf F}_{\rm st}(t) \ra = 0 \\
&&\la F_{\rm st }^{\alpha }(t) F_{\rm st}^{\beta }(t^{\prime}) =  \Gamma \delta_{\alpha \beta} \delta(t-t^{\prime}), \qquad \Gamma = 2k_BT \gamma ,  \nonumber \eeq
where $\alpha, \beta =x,y,z$, $k_B$ is the Boltzmann constant and $T$ is the temperature. For small particles the inertial effects associated with the term $m\dot{\bf v}$ are negligible, which yields the overdamped equation of motion: 
\bel{overd}
{\bf v} =\dot{\bf r} = \gamma^{-1} {\bf F}_{\rm b} + \gamma^{-1} {\bf F}_{\rm  st}.
\ee
The above Langevin equation deals with the random variable -- the particle coordinate ${\bf r}$ and the role of stochastic force plays $\gamma^{-1} {\bf F}_{\rm  st}$. Correspondingly the coefficient $\Gamma$ in the FDT relation is to be substituted by $\Gamma^{\prime} = \Gamma/\gamma^2 =2k_BT/\gamma$.    The according distribution function $P({\bf r},t| {\bf r}_0, 0)$ gives the probability of particle to reside at point ${\bf r}$ at time $t$, provided that at the initial time $t=0$ is was located at ${\bf r}_0$. It  obeys the Fokker-Planck equation \cite{krapbook,vanKampen:1992}: 

\bel{FPE}
\frac{\partial P}{\partial t} = - \nabla_{\bf r} \cdot {\bf U} P  + D \Delta_{\bf r}  P, \ee
where ${\bf U } = \gamma^{-1} {\bf F}_{\rm b}$ is the velocity of a steady particle motion under the action of the force ${\bf F}_{\rm b}$ and 
$$D= \frac{\Gamma^{\prime}}{2} =  \frac{k_BT}{\gamma} = \frac{k_BT}{6\pi \eta R}$$ 
is the diffusion coefficient. 

Now we apply the above analysis for two particles, of mass $m_1$ and $m_2$ and corresponding radii, $R_1$ and $R_2$. The according overdamped Langevin equations read:

\beel{2Lang} 
\dot{\bf r}_1 \eq \gamma_1^{-1} {\bf F}_{\rm b, 1} + \gamma_1^{-1} {\bf F}_{\rm  st,1}, \\ 
\dot{\bf r}_2 \eq \gamma_2^{-1} {\bf F}_{\rm b, 2} + \gamma_2^{-1} {\bf F}_{\rm  st,2}, \nonumber \eeq
where as previously, $\gamma_{1/2}= 6 \pi \eta R_{1/2}$ and the combined forces ${\bf F}_{\rm b , 1/2}$ and stochastic forces ${\bf F}_{\rm st, 1/2}$ obey the same relations as above, with the according values of the particles parameters.  We assume that these stochastic forces are not correlated, $\la {\bf F}_{\rm b , 1} (t){\bf F}_{\rm b , 2} (t^{\prime})\ra =0$. 

From Eqs. \eqref{2Lang} we obtain the Langevin equation for the inter-particle distance ${\bf r}_{12}$
\bel{Lan12}
\dot{\bf r}_{12} = {\bf U}_{12} + {\bf F}_{\rm st , 12}, \ee
where 
\bel{Ust}
{\bf U}_{12} =  \gamma_1^{-1} {\bf F}_{\rm b, 1} -  \gamma_2^{-1} {\bf F}_{\rm b, 2} = {\bf U}_{1}- {\bf U}_{2} \ee
is the relative steady velocity with the stochastic force 

\beel{FDT12}
&& \la  {\bf F}_{\rm st , 12} \ra =  0 \\ 
&& \la {F}_{\rm st , 12}^{\alpha}  (t) {F}_{\rm st , 12}^{\beta }  (t^{\prime})  \ra =\Gamma_{12} \delta_{\alpha \beta} \delta(t-t^{\prime}) \nonumber \\ 
&&\Gamma_{12} = \frac{2k_BT}{\gamma_1} + \frac{2k_BT}{\gamma_2} = 2(D_1+D_2).  \nonumber \eeq 
Similarly, as for the one-particle case we can write the Fokker-Planck equation for the probability density $P_{12}({\bf r}_{12},t| {\bf r}_{12,0}, 0)$ of the inter-particle separation ${\bf r}_{12}$: 
\bel{FP2}
\frac{\partial P_{12}}{\partial t} = - \nabla_{ {\bf r}_{12}} \cdot {\bf U}_{12} P_{12}  + D_{12} \Delta_{{\bf r}_{12}}  P_{12}, \ee
where $D_{12}= \frac12 \Gamma_{12} = D_1+D_2$ is the relative  diffusion coefficient for the pair of particles. 
Note that this equation has been obtained neglecting the hydrodynamic interactions between the particles. 

Now we will use Eq. \eqref{FP2} to find the aggregation rate. First we choose the direction of the gravity along $OZ$ axis and write ${\bf U}_{12} $ as ${\bf U}_{12}  =(2/9) \Delta \rho g (R_1^2-R_2^2) {\bf k} = v_{12} {\bf k}$ with $\Delta \rho = \rho_p -\rho_f$ and ${\bf k}$ being the unit vector in the $z$-direction. Next we will consider the  flux of particles of radius $R_2$ (particles of sort ``2'') on the surface of particles of radius $R_1$  (particles of sort ``1''). Let a particle of sort 1 is located at the origin and $P({\bf r}_{12},t| {\bf r}_{12,0}, 0)$ gives the probability of the location of particles of sort 2 at the distance ${\bf r}_{12}$ from the center of particles of sort 1 at time $t$. It is convenient to work with concentration $n_2({\bf r},t)$ of particles of sort 2, which is simply proportional to the probability density. To find the aggregation rate one need to compute the steady state current, of  the particles of sort 2 on the surface of the particle of sort 1 at the origin \cite{krapbook,Ven_and_Mason_1977, MELIK198484}. We assume that once the particles touch each other at $|{\bf r}_{12}|=R_1+R_2$ a joint aggregate is instantaneously formed and particles of sort 2 disappears. This implies zero concentration of particles 2 at the surface $r_{12}=(R_1+R_2)$. At the very large distance from the particle of sort 1 the concentration of particles of sort 2 equals to $n_{2, \infty}$. Then Eq. \eqref{FP2} for may be recast in the steady-state into the form,

\beel{n}
&&D \Delta n - v \frac{\del n}{\del z} =0 \\
&& n(r=R) =0 \nonumber \\
&& n(r \to \infty) =n_{\infty}.  \nonumber 
\eeq
To avoid the notation cluttering we skip here all the subscripts (i.e. $D=D_{12}$, $n=n_2$, $v=v_{12}$, $n_{\infty}=n_{2, \infty}$ and ${\bf r}={\bf r}_{12}$) and use the shorthand notation $R=R_1+R_2$. Next we introduce the dimensionless variables, ${\bf r} \to R {\bf r} $ and $n \to n_{\infty} n$, keeping for simplicity the same notations as for the dimensional quantities.  In the spherical coordinates with the $OZ$ axis along the gravity (i.e. along the velocity ${\bf U}_{12}$)  the dimensionless equation then reads
\beel{ndiml}
&&\Delta n - 2 \mu \del_z n =0, 
\label{delz} \\ 
&& \del_z n  = \cos \theta \frac{\del n}{ \del r} - \frac{\sin \theta}{r}  \frac{\del n}{\del \theta}, \nonumber \eeq
where the constant 
\bel{mudef}
\mu = \frac{vR}{2D} 
\ee
corresponds to the Peclet number.  Using additionally the substitute,  
$$
n = u(r,\theta) e^{\mu z }+1, 
$$
we recast the above equation into the form, 
\beel{u}
&&\Delta u - \mu^2 u =0 \\
\label{ubc1} && u(r \to \infty)= 0 \\
\label{ubc2} && u (r=1, \theta)= - e^{-\mu \cos \theta} .\eeq
where we choose  the spherical coordinates with $OZ$ axis along the velocity $U_{12}$ and use the symmetry of the system with respect to the azimuthal angle $\phi$. 

The separation of variables $u(r,\theta)= {\cal R}(r) \Theta (\theta)$, yields two equations, 
\begin{eqnarray}
\label{Bessel}
r^2 \frac{\partial^2 {\cal R} }{\partial r^2} + 2r\frac{\partial {\cal R} }{\partial r} - \left(\mu^2 r^2 + n(n+1) \right) {\cal R} = 0\\
\label{LegEq}
\Delta_{\theta \phi} \Theta +n(n+1) \Theta =0, 
\end{eqnarray}
where $\Delta_{\theta \phi}$ is the angular part of the Laplacian and $n=0,1, \ldots$. The solution of the first Eq. \eqref{Bessel} is ${\cal R}(r) = K_{n+\frac{1}{2}}(\mu r) /\sqrt{r}$, which the modified Bessel function of the second kind divided by the square root of $r$. The solution of the second Eq. \eqref{LegEq} are the Legendre polynomials, $\Theta (\theta)  = P_n(cos\theta)$. Hence the general solution to Eq. \eqref{u} reads, 

\begin{eqnarray}
\label{ugen} u(r, \theta) = \frac{1}{\sqrt{r}}\sum^{\infty}_{n=0} A_n P_n(cos \theta) K_{n+\frac{1}{2}}(\mu r), 
\end{eqnarray}
where the coefficients $A_n$ follow from the boundary conditions, Eqs. \eqref{ubc1} and \eqref{ubc2} \cite{tikhonov_urmat}. Expanding the boundary condition \eqref{ubc2} in terms of the Legendre polynomials, and using $r=1$ in Eq. \eqref{ugen} we find 

\beel{An}
A_n \eq  -\frac{ (2n+1) }{2\,K_{n+\frac{1}{2}}(\mu )} F_n \\
\label{Fn} F_n\eq \int_{-1}^1 e^{-\mu x}   P_n(x) dx .   \eeq
The radial component of the flux reads, 
\bel{flux}
J_r= -D \frac{\del n}{\del r}+ v\, {\bf k} \cdot {\bf e}_r n = 
-D\frac{\del n}{\del r} + v \cos \theta n,  \ee
where ${\bf e}_r$ is a unit vector in the radial direction.  On the boundary $r=R$ we obtain  the  flux 

\beel{flux2}
J_r \eq -\frac{n_{\infty} D}{R}  \left[ \frac{1}{2}-\mu \cos(\theta) \right.\\ 
 &-& \left. \mu e^{\mu \cos(\theta)}\, \sum^{\infty}_{n=0} \frac{(2n+1)K^{\prime}_{n+\frac{1}{2}}(\mu)}{2K_{n+\frac{1}{2}}(\mu)}\, F_n P_n(\cos \theta)  \right] ,  \nonumber \eeq 
where $K^{\prime}_{n+\frac{1}{2}}$ denotes the derivative of the function. The coagulation rate coefficient may be obtained integrating the radial flux over the surface of two particles contact, that is,  over the surface of radius $r=R$. Hence we obtain:
\beel{K12}
K \eq \frac{1}{n_{\infty}} \int_0^{2\pi} d \phi \int_0^{\pi} R^2 \sin \theta d \theta (-J_r) H(-J_r) \\
 \eq 2\pi DR \left[ 1-\mu \sum^{\infty}_{n=0} \frac{(2n+1)K^{\prime}_{n+\frac{1}{2}}(\mu)}{2K_{n+\frac{1}{2}}(\mu)}\, F_n C_n  \right], \nonumber \eeq
where 
\bel{Cn}
C_n = \int_{0}^{\pi}  P_n(\cos \theta ) e^{\mu \cos \theta } H \left[-J_r(\cos \theta) \right] \sin \theta  d \theta .\ee
The unit Heaviside step function $H(x)$ in the integrand of Eqs. \eqref{K12} and \eqref{Cn} guarantees that only the radial  component of the flux directed towards the particle center contributes to the coagulation rate $K$. Using the relation $xK^{\prime}_{1/2}(x)/ K_{1/2}(x) = -(x+1/2)$, one can easily check that in the limit of no advection, $\mu=0$, the standard Smoluchowski result $K=4\pi RD $ is obtained. 

The difficulties in the evaluating the coefficients $C_n$ are related to the factor $H(-J_r)$ in the integrand. Our numerical analysis has shown, that for  small Peclet numbers $\mu < 0.5$, when the diffusion dominate,  the out-flux is lacking, that is $J_r <0$ for all angles $\theta$.  This allows to skip the factor $H(-J_r)$ in Eq. \eqref{Cn} for small $\mu$. Moreover, for small $\mu$ one can also use in Eqs. \eqref{Fn} and \eqref{Cn} for $F_n$ and $C_n$ the Taylor expansion, $e^{\pm \mu x} = 1\pm \mu x +(\mu x)^2/2 +\ldots$   This yields the following analytic expression for $K$ at small $\mu$:

\bel{Ksmu}
K = 4\pi DR (1+ \mu - \frac13 \mu^2 +\ldots). \ee 
For very large $\mu$, when the advection dominates, one can neglect the diffusive motion and approximate the flux by $J= v {\bf k} n_{\infty} $ with $J_r= v n_{\infty} \cos \theta$. Eq. \eqref{K12} then yields:

\beel{Kmlar}
K \eq  2 \pi \int_0^{\pi} R^2 v \, \sin \theta  \cos \theta H(-\cos \theta)  d \theta = \pi R^2 v \nonumber \\ 
\eq 2\pi RD \mu.   \eeq
It is worth to find an extrapolating expression, which could describe not only the limiting cases $\mu \to 0$ and $\mu \to \infty$, but the whole range of $\mu$. This may be done by applying  the rational approximation which correctrly reproduces the limiting cases and hopefully provides an acceptable accuracy for the intermediate values of $\mu$. 

It is worth noting that the chosen type of rational approximation may turn out to be non-optimal from the point of view of the accuracy of approximation of the desired function, however, we consider the goal of this work, first of all, to be the correct qualitative correspondence of the constructed approximant to the asymptotics of the original kernel at zero and at infinity. Construction of such approximations is often necessary for many problems of mathematical physics \cite{andrianov2021}. Hence, in our work we expoit a standard approach \cite{Brezinski}. Let us compose an approximation as following
\begin{eqnarray}
{\cal F}(\mu) = \frac{A_0+A_1\mu+A_2\mu^2}{1 + B\mu}.
\end{eqnarray}
If one divides numerator and denominator by $\mu$ and assumes that $\mu \rightarrow \infty$ then for \eqref{Kmlar}  it is necessary 
\begin{eqnarray}
\frac{A_2}{B} = \frac{1}{2}.
\end{eqnarray}
In order to obtain other equations making possible to uniquely determine the coefficients, we expand the denominator with Taylor series up to a quadratic term. Then, with the corresponding powers and in accordance with \eqref{Ksmu}, we obtain the following relations:
\begin{eqnarray}
A_0 = 1\\
A_1 - A_0 B = 1\\
A_2 - B A_1 + A_0 B^2 = -\frac{1}{3}
\end{eqnarray}
Finally, we obtain the corresponding coefficients and substitute them itto the coagulation kernel:
\beel{Pade}
&&K= 4\pi DR  {\cal F} (\mu) \\ 
&&{\cal F} (\mu) = \left( \frac{1+ \frac{5}{3} \mu + \frac13 \mu^2}{1+ \frac23 \mu} \right). \nonumber  \eeq
It is also worth to write the coagulation kernel in the initial dimensional notations:
\beel{Kijfin}
K_{ij} \eq 4 \pi D_{ij} (R_i+R_j) {\cal F} (\mu_{ij}) \\  
\mu_{ij} \eq \frac{|v_{i}-v_{j}| (R_i+R_j)}{2D_{ij} }, \eeq
where 
\beel{DV}
&&D_{ij} = (D_i +D_j) \\
&&|v_{i}-v_{j}| = \frac29|\Delta \rho| g (R_i^2-R_j^2),
\eeq
are the mutual diffusion coefficient and  the Stoksian relative velocity of the sedimenting particles, with  $D_l =k_BT/(6 \pi \eta R_l)$, $l=i,j$. The function ${\cal F} (x)$ is defined in Eq. \eqref{Pade}. 

In the derivation  of the analytical expression \eqref{Kijfin} we neglect the hydrodynamic interactions between aggregating particles. This is twofold -- firstly, it impacts the part associated with the advection, which is accounted by the collision efficiency factor $ E(R_i, R_j)$ in Eq. \eqref{KE}. Secondly, the part associated with the mutual diffusion of two particles. To quantify the hydrodynamic correction  a solution of hydrodynamic equations for two spheres in fluid is needed. This is challenging even numerically. While the correcting factor $ E(R_i, R_j)$ for the ballistic collisions has been reported in a form of multiple tables, the analysis of the mutual diffusion coefficient has been done analytically, see e.g.  \cite{MELIK198484,Reed_1980,Ven_and_Mason_1977}. Unfortunately, all these studies are based on the expansions of the hydrodynamic forces for large inter-particle separation; analytical expressions for the forces for small distances are presently lacking. However, the most important is the impact of the hydrodynamic forces for the close location of particles,  just before a  collision. Hence derivation of reliable analytical estimates seems hardly possible. Although the derivation of the according hydrodynamic corrections is challenging, it may be worth to have,  at least a crude analytical estimate of this effect on the agglomeration kinetics. 

We propose the following phemenological modification of \eqref{Kijfin} with the hydrodynamic corrections,
\bel{Kijfin1}
K_{ij}^{\rm (h)}  = 4 \pi D_{ij} (R_i+R_j) \left( \frac{1+ \mu_{ij} +\mu_{ij}^3 E_{ij}}{1+ \mu_{ij}^2} \right) , \ee
where  $E_{ij}= E(R_i, R_j)$ is the collision efficiency factor \cite{Pinsky}. For large $\mu_{ij} \gg 1 $, when the diffusion is negligible,  the above equation will provide the kinetic rates \eqref{KE}. At the same time for $\mu_{ij} \ll 1$, the rates \eqref{Kijfin1} will coincide with  the rates of \eqref{Kijfin} up to the terms of the order of ${\cal O}(\mu_{ij}^2)$. Note that Eq.  \eqref{Kijfin1} does not account for the hydrodynamic corrections for the mutual diffusion coefficients, due to the lack of reliable analytical expressions for the near-field hydrodynamic forces. 

We expect that the above expression \eqref{Kijfin1} will be especially useful for the describing the aggregation kinetics for close-size particles, when $R_i\approx R_j$, as the standard relation  \eqref{KE} predicts vanishing aggregation rates, which is wrong. For small density distinction $\Delta \rho$, the range of the radii difference, $R_i^2 - R_j^2$, resulting in small relative velocities $|v_i-v_j| \to 0$ may be rather large, resulting in the inadequate modelling of a significant part of the aggregation reactions.

\section{Numerical simulations}

To check the accuracy of our theory for the aggregation rate coefficients when both diffusion and advection are present, we perform the numerical solution of Eq. \eqref{n} for different Peclet numbers $\mu = vR/(2D)$.  We use the implicit finite difference method rearranging the equation operators for the radial and angle components. Namely, prior to the discretization the following transformation have been done for the components of the Laplacian:

\beel{num_op}
\Delta_r n  &= &\frac{1}{r^2}\frac{\partial}{\partial r} \left( \frac{r^2}{g}  \frac{\partial (g n )}{\partial r} \right) \\
\Delta_{\theta} n &= & \frac{1}{r^2\sin \theta}\frac{\partial}{\partial \theta} \left( \frac{ \sin \theta}{g} \frac{\partial (g n)}{\partial \theta}  \right)  \\ 
g &= & e^{-2 \mu r \cos \theta}. 
\eeq
This transformation is needed to stabilize the numerical error occurring at high Péclet numbers $\mu$. Incorporating of the exponential factor $g$ results in the terms like $\mu \,  (\cos \theta_j-\cos \theta_{j \pm 1/2} ) $ and $\mu \, dr$ (the subscripts  refer to the numbers  of the grid points) which can reduce the destabilizing effect of large $\mu$ by applying smaller discretization steps \cite{Vabishchevich}. 

All in all, we obtain the finite-difference scheme corresponding to equations \eqref{num1} -- \eqref{num3}, where $i=\{1, \ldots, N-2\}$, $j=\{0, \ldots, M-1\}$, $h_r$ is a grid step for radial variable and $h_\theta$ is a  grid step for angular variable. For $i=0$, $i=N-1$ we use the original boundary conditions \eqref{n}.
\begin{widetext}
\begin{eqnarray}
 \Lambda_r = \frac{1}{r^2_i}\frac{1}{h_r} \left(r_i r_{i+1}\frac{n_{i+1,j}e^{-2\mu \cos \theta_j (r_{i+1}-r_{i+0.5})}-n_{i,j}e^{-2\mu \cos \theta_j (r_{i}-r_{i+0.5})}}{h_r} \right. \notag \\
\label{num1}
\left. -r_i r_{i-1}\frac{n_{i,j}e^{-2\mu \cos \theta_j (r_{i}-r_{i-0.5})}-n_{i-1,j}e^{-2\mu \cos \theta_j (r_{i-1}-r_{i-0.5})}}{h_r}\right), \\
 \Lambda_\theta = \frac{1}{r_i^2 \sin \theta_j}\frac{1}{h_\theta}\left(\sin \theta_{j+0.5}\frac{e^{-2\mu r_i (\cos \theta_{j+1}-\cos \theta_{j+0.5})}n_{i, j+1}-e^{-2\mu r_i (\cos \theta_{j}-\cos \theta_{j+0.5})}n_{i, j}}{h_\theta} \right. \notag \\
 \label{num2}
\left. -\sin \theta_{j-0.5}\frac{e^{-2\mu r_i (\cos \theta_{j}-\cos \theta_{j-0.5})}n_{i, j}-e^{-2\mu r_i (\cos \theta_{j-1}-\cos \theta_{j-0.5})}n_{i, j-1}}{h_\theta}\right), \\
\label{num3}
 \Lambda_r + \Lambda_\theta = 0,
\end{eqnarray}
\end{widetext}
If we split the grid over $\theta$ in such a way that $\theta_{-0.5} = 0$ and $\theta_{M-1+0.5} = \pi$ then due to the multipliers $\sin \theta_{j-0.5}$ and $\sin \theta_{j+0.5}$ the angular opetator stays within the borders of the computational domain over $n$ without the necessary definition of the boundary conditions for the angular variable. 

Finally, we construct a system of linear equations from the resulting scheme with a sparse matrix of size $NM \times NM$, where $N$ is the number of points in $r$, $M$ is the number of points for $\theta$. The solution of this linear system  gives us the numerical solution of the equation \eqref{n}. We use the standard direct solver from umfpack \cite{umfpack} package through the \texttt{scipy.sparse.linalg} interface in the \texttt{Python3} programming language for solving these linear systems. 

\begin{figure}
\centering
\includegraphics[width=8.5cm]{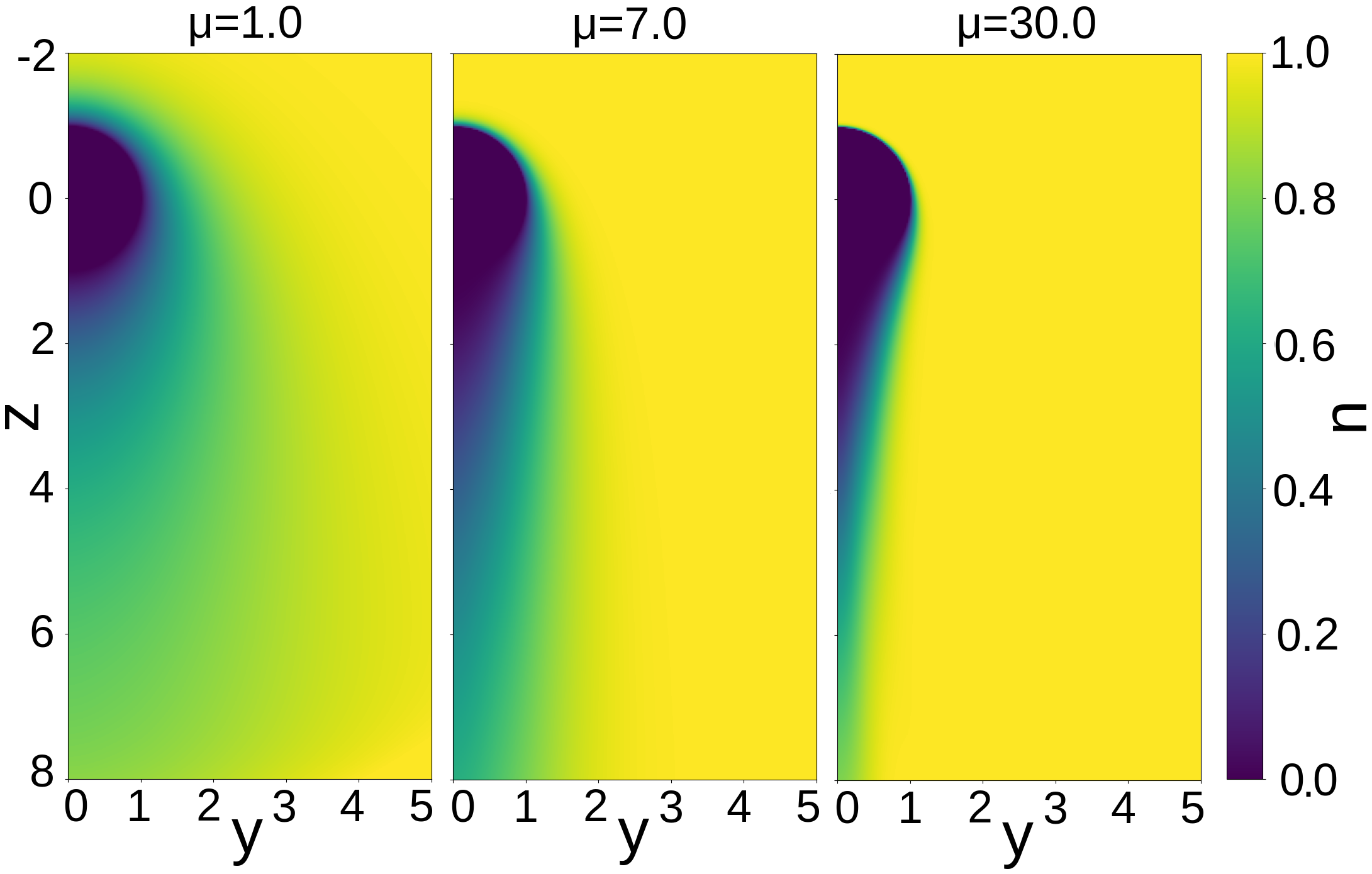} 
\caption{The numerical solution of the steady-state convection-diffusion equation \eqref{n} for different Péclet numbers $\mu = vR/(2D)$.}
\label{cartesian} 
\end{figure}

\begin{figure}
\centering
\includegraphics[width=8.5cm]{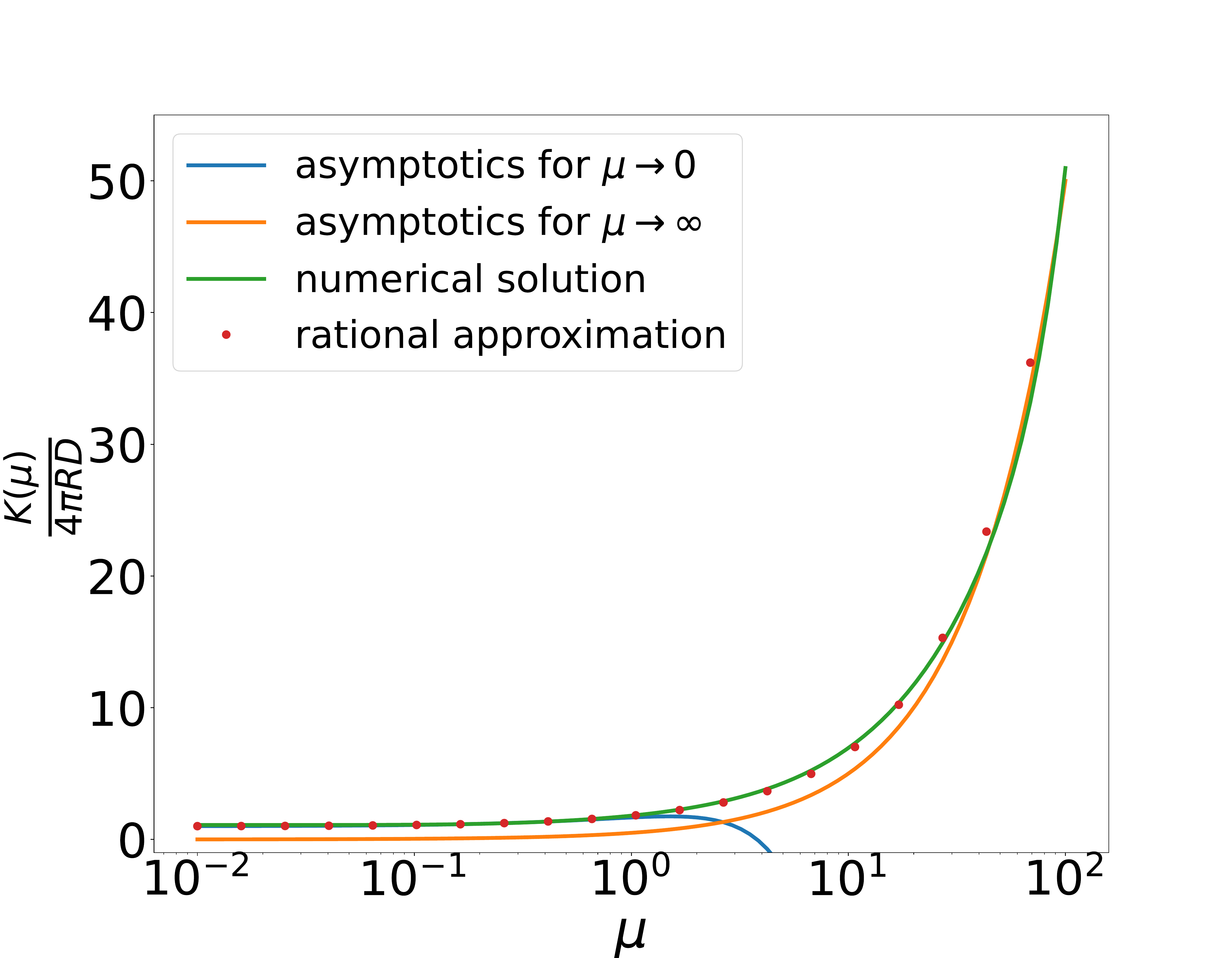} 
\caption{ The reduced aggregation  rate $K/(4\pi RD)$ as the function of the Peclet number $\mu = vR/(2D)$. The result obtained by the direct numerical solution of the advection-diffusion equation \eqref{n} (solid green line) is very close to the analytical rational approximant \eqref{Pade} for the whole range of $\mu$. Also shown the asymptotic  \eqref{Ksmu} for $\mu \to 0$ (solid blue line) and \eqref{Kmlar} for $\mu \to \infty$ (solid brown line). The analytical expression approximates the numerical results extremely well. } 
\label{appro_pic} 
\end{figure}

We investigate numerically the following range of  Peclet number,  $ \mu \in [0.01, 100.0]$, using the grid of 100 points for the angular coordinate $\theta$ and 500 points for the radial coordinate $r$. The error of the numerical scheme,  based on the Chebyshev norm, was less than $0.01$ in all simulations. 
The result for the concentration distribution $n$ is depicted in Fig.  \ref{cartesian}
for three different Peclet numbers. It illustrates the disappearance of a diffusive ``halo'' for large $\mu$, existing for smaller $\mu$, in accordance with the intuitive expectations, that for large Peclet numbers diffusion does not contribute. 
In Fig. \ref{appro_pic} the results for the aggregation rate $K$, obtained by the  direct numerical solution of Eq. \eqref{n} is compared with the analytical result \eqref{Pade}. The figure demonstrate an astonishing accuracy of the analytical approximation for the whole range of the Peclet numbers, covering four orders of magnitude of $\mu$.  

\section{Conclusion} 
We investigate numerically and theoretically the agglomeration kinetics of sedimenting particle, based on the advection-diffusion equation. Both transport mechanisms -- the  diffusion and advection are important for aggregation in two cases. Firstly,  for small particles, like e.g. primary ash or soot particles in the atmosphere. Secondly, for large particles, of equal or close size -- such particles sediment with  almost the same velocity, making the advection mechanism of aggregation faint, comparable or even weaker than the diffusional one. Hence, it is desirable to have the aggregation rates for such systems, where the relative importance of the above mechanisms (say diffusional) varies from the prevailing one to negligible. In our analysis we do not consider explicitly the hydrodynamic interactions between  agglomerating particles: For  dominating advection we use the phenomenological factor of collision efficiency. For dominating diffusion we neglect the hydrodynamic forces due to the lack of reliable expressions for these forces at  close inter-particle distances.      

For small Peclet numbers, which quantify the relative importance of the diffusion and advection, we derive an analytical expression for the aggregation rates, in the form of expansion in Peclet numbers. For very  large Peclet numbers we use the according expression for a purely ballistic relative motion of the particles. Using these results we construct the rational approximant for the whole range of Peclet numbers. We also solve the  advection-diffusion equation numerically, applying the technique which increases the accuracy of the numerical scheme. The  results of the numerical simulations are in an excellent agreement with the analytical theory for the all studied Peclet numbers, varying by four orders of magnitude. 

\section*{Acknowlegement}  

Work of R. Zagidullin, A. Smirnov and N. Brilliantov (analytical deriviation of the equations, problem setting and numerical calculations) has been supported  by the grant of  Russian Science Foundation No. 21-11-00363,
https://rscf.ru/project/21-11-00363/. Work of S. Matveev (construction of the rational approximation) has been supported by the grant of the Russian Science Foundation 19-11-00338.

\bibliography{lib}

\end{document}